\documentclass{Interspeech}

\usepackage{cite}
\usepackage{amsmath,amssymb,amsfonts}
\usepackage{algorithmic}
\usepackage{graphicx}
\usepackage{textcomp}
\usepackage{xcolor}
\usepackage{hyperref}
\usepackage{siunitx}
\usepackage{graphicx}
\usepackage{booktabs}
\usepackage{makecell}
\usepackage{multirow} 
\usepackage{caption}
\usepackage{mathrsfs}
\usepackage{bm}       
\usepackage{subcaption}  
\usepackage{xurl}
\usepackage{hyperref}

\interspeechcameraready

\title{A Perception-Based L2 Speech Intelligibility Indicator: \\ Leveraging a Rater’s Shadowing and Sequence-to-sequence Voice Conversion}

\author[affiliation={1}]{Haopeng}{Geng}
\author[affiliation={1}]{Daisuke}{Saito}
\author[affiliation={1}]{Nobuaki}{Minematsu}

\affiliation{Graduate School of Engineering}{The University of Tokyo}{Japan}
\email{\{kevingenghaopeng, dsk\_saito, mine\}@gavo.t.u-tokyo.ac.jp}

\usepackage{comment}

\begin{document}

\maketitle

\begin{abstract}
    

Evaluating L2 speech intelligibility is crucial for effective computer-assisted language learning (CALL). Conventional ASR-based methods often focus on native-likeness, which may fail to capture the actual intelligibility perceived by human listeners. In contrast, our work introduces a novel, perception-based L2 speech intelligibility indicator that leverages a native rater’s shadowing data within a sequence-to-sequence (seq2seq) voice conversion framework. By integrating an alignment mechanism and acoustic feature reconstruction, our approach simulates the auditory perception of native listeners, identifying segments in L2 speech that are likely to cause comprehension difficulties. Both objective and subjective evaluations indicate that our method aligns more closely with native judgments than traditional ASR-based metrics, offering a promising new direction for CALL systems in a global, multilingual contexts.

\end{abstract}

\section{Introduction}
\label{intro}
\begin{figure*}[t]
    \centering
    \begin{subfigure}[b]{0.48\textwidth}
        \centering
        \includegraphics[width=\textwidth]{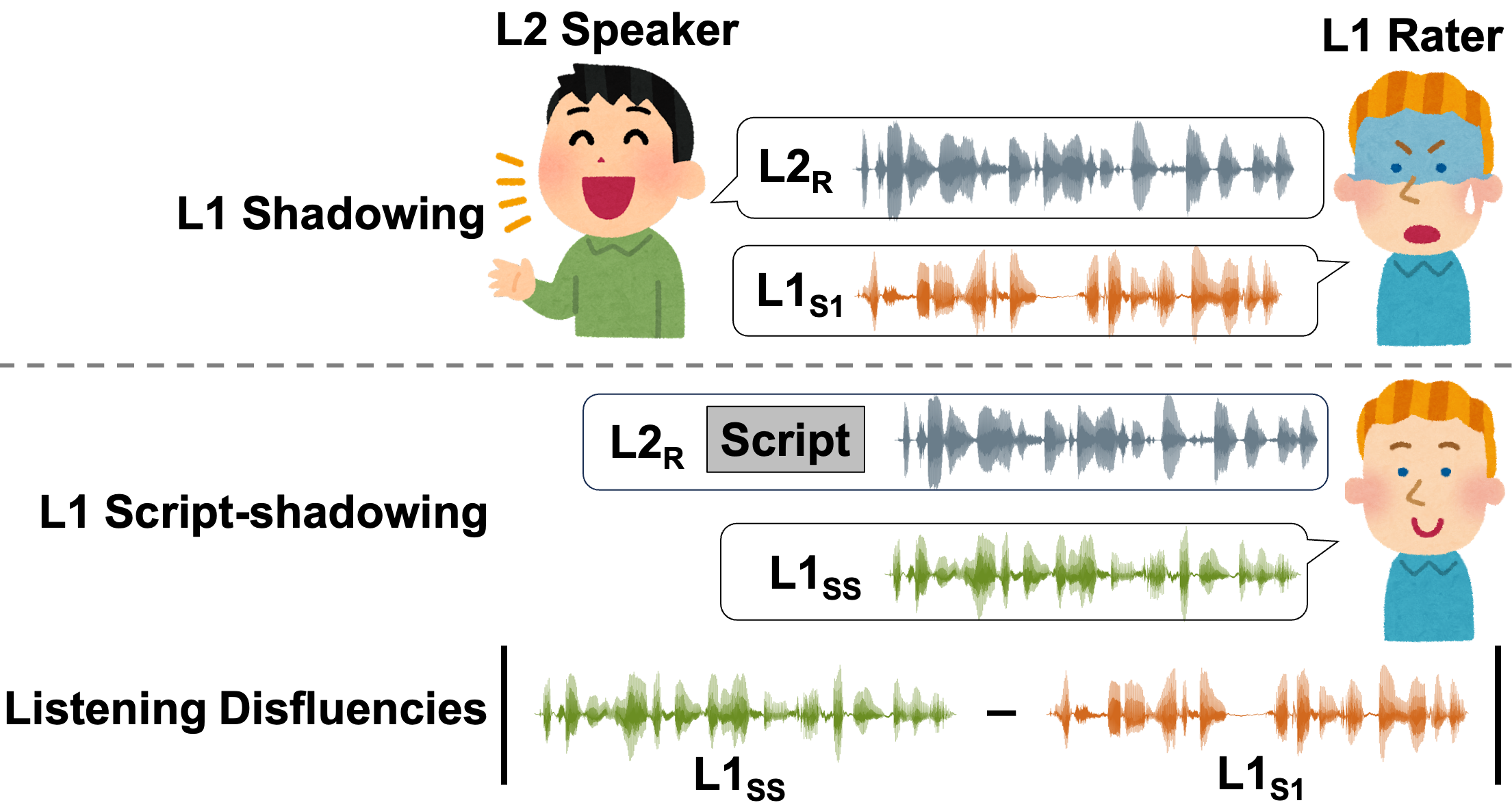}
        \caption{L1 rater shadowing for L2 speech intelligibility assessment.}
        \label{fig:l1shadowing}
    \end{subfigure}
    \hfill
    \begin{subfigure}[b]{0.48\textwidth}
        \centering
        \includegraphics[width=\textwidth]{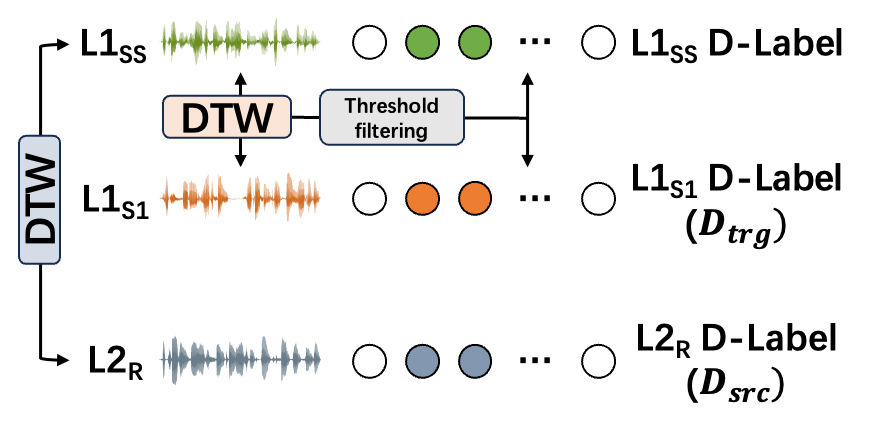}
        \caption{Mapping disfluency label from a rater's shadowing to L2 speech.}
        \label{fig:Dlabel}
    \end{subfigure}

    \caption{An overview of the L1 rater’s shadowing technique for immediate L2 speech intelligibility assessment. The proposed approach is designed to identify unintelligible segments in the L2 speaker’s read-aloud utterances ($L2_{R}$).}
    \label{fig:shadowing}
    \vspace{-7mm}
\end{figure*}

Non-native speakers are often encouraged to emulate native-like pronunciation; however, achieving a native-like accent can be time-consuming and may not efficiently improve speech intelligibility. 
Many studies indicate that mispronunciations or accents do not necessarily result in unintelligible speech. For example, \cite{loukina15_interspeech} found that only 16\% of word-level unintelligibility in an English spoken test can be attributed to obvious mispronunciations, and experimental evaluations in \cite{graham2024evaluating} show that even non-native accents, such as Swedish and German-accented English, yield higher recognition accuracy using OpenAI's Whisper compared to British-accented English. 

According to \cite{ethnologue_most_spoken}, approximately 1.14 billion non-native speakers account for 75\% of the global English-speaking community. However, mainstream Automatic Speech Recognition (ASR) systems are typically built using native-accented corpora or the most prevalent accents, which can lead to biased performance that often results in higher error rates and misrecognition on non-native or less common accents. Additionally, recent ASR systems are specifically designed to predict—or even over-speculate on—what speakers actually said, which conflicts with the goals of computer-assisted language learning (CALL) systems \cite{Derwing_Munro_1997, munro1995foreign}. From an educational perspective, language learners require feedback on whether their accents hinder communication with speakers from diverse backgrounds rather than simply striving for native-like pronunciation. Therefore, it is essential to develop methods that capture not only native-like pronunciation but also the actual intelligibility as perceived by real human listeners.

How can we effectively identify unintelligible parts in L2 speech? One promising method is to measure native (L1) listeners’ responses by asking them to shadow L2 utterances—where an L1 speaker immediately repeats the L2 speech with minimal delay. This approach, as explored in previous research \cite{inoue18_interspeech, lin2020shadowability, zhu2021multi}, has proven effective at revealing where listening breakdowns occur, highlighting problematic segments in the L2 speech, and providing valuable feedback for learners.

Inspired by this approach, our study aims to utilize the L1 rater's shadowing process to build an unintelligibility indicator based on the rater's perception. Our contributions are as follows:
\begin{enumerate}
    \item We propose a customized speech unintelligibility indicator that leverages native rater’s shadowing data. Unlike conventional methods that focus solely on native-like pronunciation, our approach aims to capture the perceptual cues that native listeners actually use to assess intelligibility\footnote{Samples are available at \url{https://secondtonumb.github.io/publication_demo/IS_2025/}}.
    \item Our method incorporates a state-of-the-art sequence-to-sequence (seq2seq) voice conversion framework, utilizing its alignment mechanism and acoustic feature reconstruction module to simulate the cognitive process of native listeners in identifying unintelligible segments.
    \item We introduce a multi-task learning strategy that jointly optimizes speech reconstruction and disfluency detection using auxiliary loss functions. Experimental evaluations demonstrate that our approach aligns more closely with native raters' perceptions than mainstream ASR, revealing the potential for delivering personalized feedback in CALL systems.

\end{enumerate}
\section{Research background}

\subsection{L1 rater's shadowing}
\label{l1ratershadowing}
Assessing L2 speech intelligibility requires fine-grained annotation, yet this process is challenging due to the need for expert-level phonetic knowledge for phoneme-level labeling. To address this challenge, \cite{lin2020shadowability} proposed a two-stage reverse shadowing approach to identify unintelligible segments in L2 utterances\footnote{In second language acquisition (SLA), shadowing usually involves learners repeating L1 speech to improve their listening skills. In our study, however, L1 raters shadow L2 speech—a process we refer to as reverse shadowing.}. As illustrated in Fig~\ref{fig:l1shadowing}, an L1 speaker first shadows a given L2 utterance alone (\(L1_{S1}\)), during which listening breakdowns manifest as stuttering or inarticulate speech. Subsequently, the L1 speaker performs script-shadowing (\(L1_{SS}\)), where the L2 script is displayed while the corresponding L2 utterance is played acoustically. By comparing \(L1_{S1}\) and \(L1_{SS}\) using dynamic time warping (DTW), this method effectively captures sequential disfluency patterns that indicate listening breakdowns. Prior research has demonstrated that phonetic posteriorgram (PPG)-based DTW alignment between \(L1_{S1}\) and \(L1_{SS}\) can effectively yield unintelligibility annotations at the word, syllable, and phoneme levels \cite{zhu2021multi, yue17_interspeech}.

Fig~\ref{fig:Dlabel} illustrates the process for obtaining frame-wise disfluency labels on \(L2_{R}\) using the two-stage reverse shadowing approach. First, DTW is applied between the first shadowing speech \(L1_{S1}\) and the script-shadowing speech \(L1_{SS}\) to capture listening breakdowns in \(L1_{S1}\). Then, by thresholding the edit distance between \(L1_{S1}\) and \(L1_{SS}\), a disfluency label (\(\mathbf{D}\)-Label) can be derived for both \(L1_{SS}\) and \(L1_{S1}\). Finally, another DTW alignment between \(L1_{SS}\) and \(L2_{R}\) is used to map the \(\mathbf{D}\)-Label onto \(L2_{R}\), thereby identifying the unintelligible frames.

\subsection{Seq2seq voice conversion with self supervised speech representations}
\label{seq2seqvc}
Conventional voice conversion (VC) seeks to modify non-/para-linguistic features while preserving the linguistic content of the input speech. With the advent of end-to-end architectures, recent seq2seq VC models have achieved robust mappings between sophisticated source and target features. For example, \cite{huang20i_interspeech} introduced a transformer-based VC model capable of sequentially mapping speech with variable durations. Meanwhile, limitations of autoregressive models, such as inaccurate duration prediction and repetitive artifacts, was addressed by proposing a non-autoregressive model using a conformer structure in \cite{hayashi2021non}. More recent approaches combine Monotonic Alignment Search (MAS) with joint vocoder training \cite{okamoto2023e2e, huang2023aas}, which enhances the precision of duration modeling and improves prosody synthesis by reducing alignment errors.

Advances in self-supervised learning have further enhanced VC by improving embedding features. Self-supervised speech representations (S3Rs) such as Wav2vec 2.0 \cite{wav2vec2}, HuBERT \cite{hsu2021hubert}, and WavLM \cite{chen2022wavlm} have set new benchmarks across diverse speech tasks \cite{yangspeechfound}. Consequently, self-supervised speech representation-based VC (S3R-VC) has demonstrated notable gains in both linguistic accuracy and acoustic quality, while also proving effective for processing unintelligible speech such as electrolaryngeal speech enhancement \cite{lester_EL, violeta2024electrolaryngeal}. Moreover, the combination of improved architectures and advanced embeddings has enabled novel mappings between source and target speech—such as shadowing-like speech generation \cite{geng2024APSIPA}—even when the linguistic content of the two voices is not fully parallel.

In the following sections, we describe how we adapt these recent advances in seq2seq VC, particularly the enhanced alignment module, for both speech reconstruction and disfluency detection.

\section{Proposed Methods}
In this section, we present two approaches for evaluating L2 speech intelligibility based on seq2seq VC. The first approach leverages alignment failure to mimic rater perception breakdowns, thereby identifying segments of unintelligibility. The second approach employs a multi-task learning strategy in a Seq2Seq VC model, where additional loss functions guide the model to detect unintelligibility.

\begin{figure*}[t]
    \centering
    \includegraphics[width=\linewidth,trim=7 0 7 7,clip]{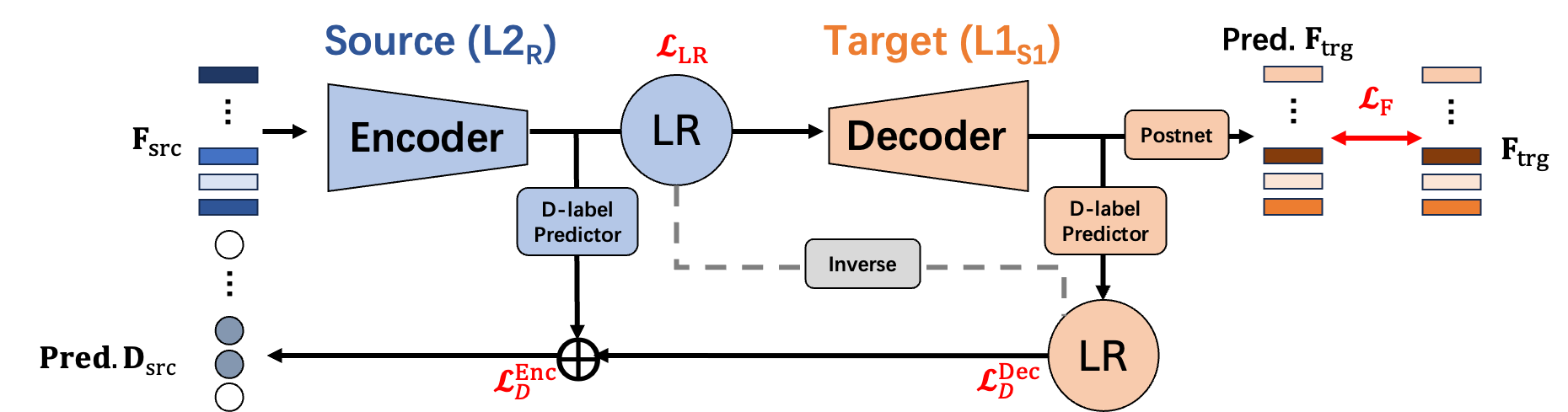}
    \captionsetup{skip=2pt}
    \caption{Proposed multi-task learning for L2 intelligibility prediction based on seq2seq VC.}
    \label{fig:Model_D_on_source}
    \vspace{-7mm}
\end{figure*}

\subsection{Using alignment breakdowns to replicate rater's perception breakdowns} 
\label{align_fail}

In parallel TTS and VC research, the MAS technique mentioned in Sec.~\ref{seq2seqvc} has been widely adopted \cite{shih2021radtts, onetts}. In this process, a soft alignment matrix \(\mathcal{A}_{\text{soft}} \in \mathbb{R}^{T_{\text{src}} \times T_{\text{trg}}}\) is generated based on the pairwise similarity between the source and target features. The Viterbi algorithm is then used to compute the hard alignment path between, denoted by \(\mathcal{A}_{\text{hard}}\). The alignment is optimized by minimizing the discrepancy between \(\mathcal{A}_{\text{soft}}\) and \(\mathcal{A}_{\text{hard}}\) using KL-divergence. Specifically, the binary loss is defined as
\begin{align}
\mathcal{L}_{\text{bin}} &= \mathcal{A}_{\text{hard}} \odot \log \mathcal{A}_{\text{soft}} \label{eq:Lbin}
\end{align}
where \(\odot\) denotes the Hadamard product and the final alignment loss is combined with the forward sum loss:
\begin{align}
\mathcal{L}_{\text{align}} &= \mathcal{L}_{\text{ForwardSum}} + \mathcal{L}_{\text{bin}} \label{eq:Lalign}
\end{align}

This idea is based on the assumption that the source and target are parallel. However, in our scenario, shadowing, a speech-to-speech process through perception, often results in speech from the rater that does not perfectly match the speech from the learner. Consequently, when aligning these semi-parallel data, segments with noticeable differences are expected to misalign. This observation motivates us to treat the alignment module's failure segments as unintelligible segments in \(L2_{R}\) speech. Given the converged non-positive matrix \(\log\mathcal{A}_{\text{soft}}\), the frame-wise alignment focus rate \(s_i\) can be calculated for each source frame \(i\) (\(1 \leq i \leq T_{\text{src}}\)) as follows:
\begin{align}
s_i = \mathbb{I}\Bigl\{ \max_{j \in \{1, \dots, T_{\text{trg}}\}} \log \mathcal{A}_{\text{soft}}^{(i,j)} < \tau \Bigr\}
\end{align}
where \(\tau\) is a predefined threshold, and \(\mathbb{I} \{\cdot\}\) is the indicator function that returns 1 if the condition is true, indicating that frame \(i\) contains segments that lead to perception failure during the rater's shadowing process.

\subsection{Multi-task learning: seq2seq VC with disfluency detection}
\label{multi_task}
Another approach aims to convert $L2_{R}$ to $L1_{S1}$ while simultaneously detecting unintelligible segments in $L2_{R}$, is illustrated in Fig~\ref{fig:Model_D_on_source}. In typical seq2seq VC models, let 
$
\mathbf{F}_{\text{src}} \in \mathbb{R}^{T_{\text{src}} \times d}$ and $ \quad \mathbf{F}_{\text{trg}} \in \mathbb{R}^{T_{\text{trg}} \times d}$
denote the continuous features of the source and target utterances, respectively. The transformation process in the VC model is defined as follows:
\begin{align}
\mathbf{h} &= \text{Enc}(\mathbf{F}_{\text{src}}) \label{eq:enc} \\
\mathbf{h'} &= \text{LR}(\mathbf{h}) \label{eq:lr} \\
\mathbf{z} &= \text{Dec}(\mathbf{h'}) \label{eq:dec} \\
\mathbf{F}_{\text{trg}} &= \text{Postnet}(\mathbf{z}) \label{eq:postnet}
\end{align}
where the length regulator (LR) maps the source hidden representation 
\(\mathbf{h} \in \mathbb{R}^{T_{\text{src}} \times h}\) to the target representations \(\mathbf{h'}\) and \(\mathbf{z} \in \mathbb{R}^{T_{\text{trg}} \times h}\). 

Directly capturing the positional information of unintelligibility from $
\mathbf{F}_{\text{src}}$ is challenging because the source speech, although accented, remains fluent. To tackle this, we propose guiding the model by incorporating additional modules aimed at detecting disfluency on the target side, since the disfluency information is explicit in the rater's shadowing voice. Specifically, we integrate two disfluency label predictors (DLPs), each implemented using a 5-layer CNN, which are positioned after the encoder and decoder, respectively. By leveraging the hidden outputs from both the encoder and decoder, the model can effectively capture disfluency information. This information is then projected back to the source side via an inverse length regulation, serving as a proxy for the source’s unintelligibility. The disfluency position information is computed as follows:
\begin{align}
\mathbf{D}_{\text{src}} &= \text{Linear}\left( \text{DLP}(\mathbf{h}) \oplus \text{LR}^{-1}(\text{DLP}'(\mathbf{z})) \right)
\end{align}
where \(\text{LR}^{-1}\) is the inverse length regulation from target to source. Finally, incorporating our proposed DLPs, whose loss functions \(\mathcal{L}_{\text{D}}^{\text{Enc}}\) and \(\mathcal{L}_{\text{D}}^{\text{Dec}}\) are implemented using the focal loss \cite{focalloss}, we define the overall loss as:
\begin{align}
\mathcal{L}_{\text{VC}} &= \mathcal{L}_{\text{F}} + \mathcal{L}_{\text{LR}} + \mathcal{L}_{\text{align}}\\[1ex]
\mathcal{L}_{\text{all}} &= \lambda \left( \mathcal{L}_{\text{D}}^{\text{Enc}} + \mathcal{L}_{\text{D}}^{\text{Dec}} \right) + \mathcal{L}_{\text{VC}}
\end{align}
where \(\mathcal{L}_{\text{F}}\) is the L1 loss between the predicted and target acoustic features, and \(\mathcal{L}_{\text{LR}}\) is the loss for predicting the appropriate duration transformation referring \cite{huang2023aas}. The hyper-parameter \(\lambda\) was set to 10 to balance the training criteria. This combined loss formulation ensures balanced target feature reconstruction and D-label detection for binary unintelligibility detection.

\begin{table}[t!]
\centering
\caption{Objective evaluation of the proposed L1 perception-based intelligibility indicator }
\label{tab:objective}
\begin{tabular}{@{}lcccc@{}}
\toprule
\multirow{2}{*}{\textbf{Description}} & \multicolumn{3}{c}{\textbf{Word}} & \textbf{Frame} \\ 
\cmidrule(r){2-4}
\cmidrule(l){5-5}
       & F1$\uparrow$& Precision$\uparrow$ & Recall$\uparrow$ & Acc. $\uparrow$\\ 
\midrule
ASR-based & 28.0 & 28.8 & 27.2 & --- \\ 
\midrule
Alignment-based  & 27.2 & 23.0 & \textbf{33.3} & --- \\ 
\midrule
Multi-task: &&&&\\ 
\quad  \(\mathcal{L}_{D}^{\text{Enc}}\)   & 25.5 & 25.0 & 26.0 & 74.2 \\ 
\quad  \(\mathcal{L}_{D}^{\text{Dec}}\)   & 23.5 & 23.0 & 24.0 & 73.3 \\ 
\quad  \(\mathcal{L}_{D}^{\text{Enc}} + \mathcal{L}_{F}\)  & 28.6 & 26.1 & 31.6 & 74.3 \\ 
\quad  \(\mathcal{L}_{D}^{\text{Dec}} + \mathcal{L}_{F}\)  & 24.5 & 25.5 & 23.6 & \textbf{75.7} \\ 
\quad  \(\mathcal{L}_{D}^{\text{Enc}}+\mathcal{L}_{D}^{\text{Dec}} + \mathcal{L}_{F}\)  & \textbf{30.3} & \textbf{29.2} & 31.4 & 75.6 \\ 
\bottomrule
\end{tabular}
    \vspace{-7mm}

\end{table}
\section{Experimental Settings}

\subsection{Dataset}
\label{dataset}
We utilized the shadowing dataset described in \cite{yue17_interspeech, zhu2021multi}. Reading-aloud utterances were collected from 225 Japanese learners of English ($L2_{R}$) with varying degrees of Japanese accents. An American male English speaker (EN-A) then shadowed and script-shadowed these recordings, producing $L1_{S1}$ and $L1_{SS}$, respectively. In total, 2,695 triplets \{$L2_{R}$, $L1_{S1}$, $L1_{SS}$\} were prepared. This dataset contained 3.9 hours of valid phonation, with an average duration of 5.0 seconds. Three hundred utterances were selected as the test set, while the remaining utterances were allocated as the training and development sets.

For unintelligibility detection, we applied the method described in Sec.~\ref{l1ratershadowing} to generate frame-wise labels with 10 ms hop size. We then converted the frame-level D-labels to word-level via weighted editing path concatenation, following the approach outlined in \cite{zhu2021multi}, since word-level degradation more closely aligns with human perception.

\subsection{Feature embedding}
In our preliminary experiments, we initially attempted to use HuBERT for both the source and target phases of feature mapping. However, the alignment did not converge as expected based on our observation. Consequently, based on implementations in \cite{geng2024APSIPA}, we adopted the ninth-layer output of the HuBERT-base model on the source side and a PPG-like bottleneck feature (PPG-BNF) from \cite{liu2021any} for the target, which yielded the best alignment. The feature dimensions per frame were 768 and 144, respectively.


\section{Experimental Evaluation}
\label{exp_eva}

\subsection{Objective evaluation}

We present our objective evaluation results in Table~\ref{tab:objective}. Both frame-level and word-level metrics are considered, with the word-level evaluation computed as an F1-score based on the positional detection of unintelligible words in $L2_{R}$. The conventional approach—relying on an ASR system trained on librispeech dataset \cite{librispeech} for mispronunciation detection—serves as our baseline\footnote{ \url{https://huggingface.co/facebook/wav2vec2-large-960h-lv60-self}}. We first evaluated an alignment-based method (described in Sec.~\ref{align_fail}). Although this method achieved an improved recall rate compared to the baseline, the degradation in F1-score and precision indicates that 
this method tends to over-predict unintelligible segments, which means further development is required.

In our proposed multi-task learning framework (detailed in Session~\ref{multi_task}), we conducted an ablation study by incrementally incorporating the auxiliary loss functions $\mathcal{L}_{\text{D}}^{\text{Enc}}$, $\mathcal{L}_{\text{D}}^{\text{Dec}}$ and $\mathcal{L}_{\text{F}}$. As shown in Table~\ref{tab:objective}, predicting D-labels solely from the source features—without leveraging target speech—resulted in suboptimal performance. In particular, when all auxiliary loss functions were applied, our multi-task approach attained the highest word-level F1-score of 30.3 and the second-highest frame-level accuracy of 75.6, which is only 0.1 \% lower than the best result—demonstrating nearly equivalent accuracy to the top-performing method.


\subsection{Subjective evaluation}
\begin{figure}[t]
    \centering
        \includegraphics[width=\linewidth,trim=8 0 8 8,clip]{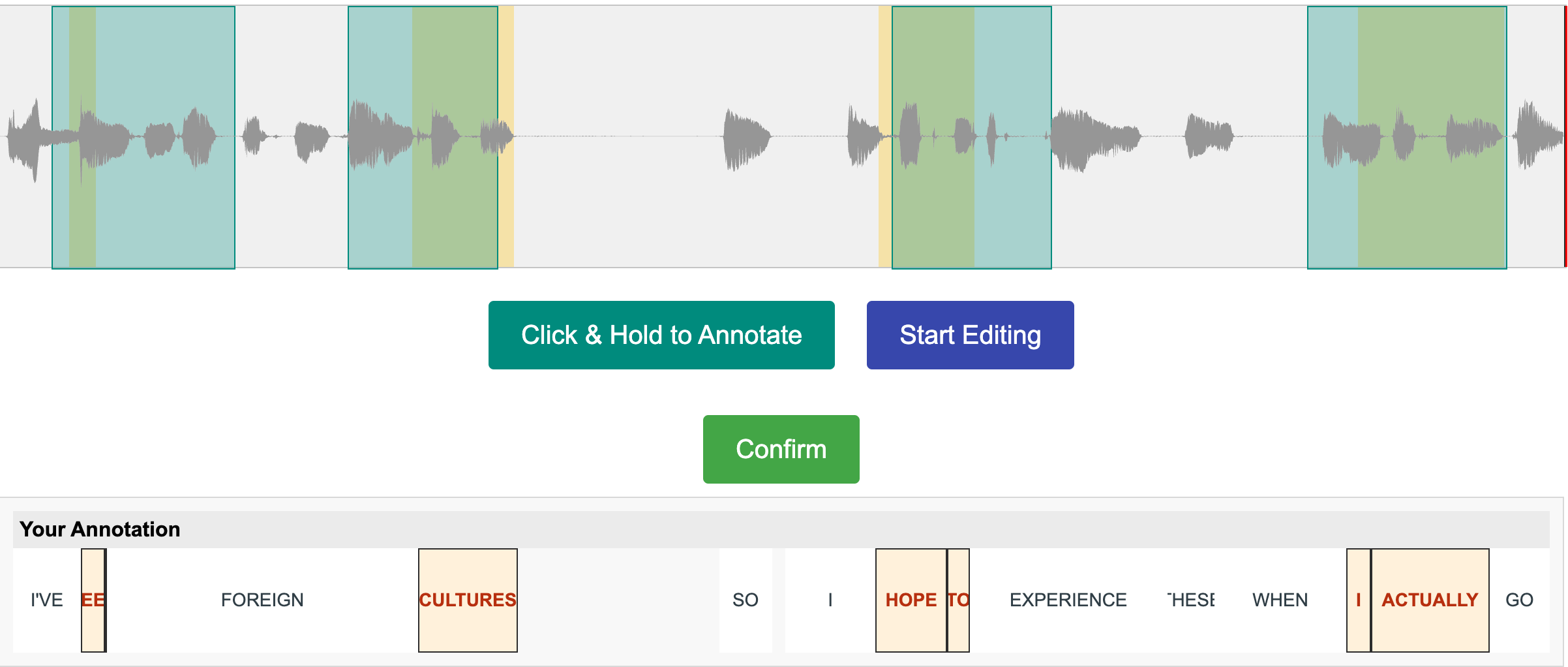}
        \caption{Overview of annotation system for unintelligibility.}
    \label{fig:annotation_sys}
    \vspace{-7mm}

\end{figure}
Objectively evaluating the perception-based assessment is challenging because precisely annotating the onset and offset of intelligibility corruption is inherently difficult. Thus, to examine whether our system can effectively simulate the rater’s cognition, we conducted a subjective evaluation. In this evaluation, both the rater and other English speakers with diverse backgrounds were asked to annotate segments of the speech that were not immediately intelligible to them. Figure~\ref{fig:annotation_sys} illustrates our annotation system\footnote{\url{https://secondtonumb.github.io/DisfluencyAnnotationExp}}, which is inspired by \cite{coulange2024exploring}.
Our participants included:
\begin{itemize}
    \item \textbf{EN-A}: The American rater mentioned in Session~\ref{dataset}\footnote{To reduce memory effects, the subjective annotations were conducted four years after the shadowing experiment and a randomly selected set of utterances were used}
    \item \textbf{EN-J}: A Japanese speaker learning English.
    \item \textbf{EN-O}: Six L2 English speakers from diverse linguistic backgrounds (excluding English and Japanese).
\end{itemize}
Participants were instructed to annotate segments of \(L2_{R}\) that they could not identify words by holding and releasing an annotation button while the utterance was playing. After the initial real-time annotation, the participants were allowed to edit mis-annotated segments. The annotated frames were then used to pinpoint the corresponding words that were unintelligible to the participants. Forced alignment between the waveform and the text sequence of \(L2_{R}\) was performed using Kaldi’s WSJ implementation.


\begin{table}[t]
  \centering
  \caption{Precision of the proposed unintelligibility indicator compared to manual annotation}
  \begin{tabular}{@{}lccc@{}}
    \toprule
    \multirow{2}{*}{\textbf{Methods}} & \multicolumn{3}{c}{\textbf{Precision}$\uparrow$}  \\
    \cmidrule(lr){2-4}
                            & EN-A & EN-J & EN-O \\
    \midrule
    2-Stage Shadowing     + DTW \\
    (Ground Truth)                 & 45.9 & 40.6 & 31.5 \\
    \midrule
    ASR-based                   & 40.5 & 32.8 & 37.6 \\
    Alignment-based               & 21.6 & 34.4 & 27.0 \\
    Multi-task                    & 43.2 & 28.1 & 31.7 \\
    \bottomrule
  \end{tabular}
  \label{tab:sub}
    \vspace{-5mm}

\end{table}
Table~\ref{tab:sub} presents a comparison between the predicted unintelligible segments and the participants' annotations. We evaluated the similarity between the manual annotations and the predictions of the proposed method in terms of precision rate, defined as the proportion of correctly identified unintelligible segments over all predicted segments. 

Among all methods, the 2-stage shadowing approach using DTW achieved the highest precision, demonstrating its effectiveness in replicating the rater’s cognition of L2 speech. Furthermore, comparing the annotations of EN-J and EN-O revealed that the PPG-DTW results more closely resemble those of the Japanese learner (EN-J). This is reasonable because Japanese learners are generally more familiar with their own accented English, whereas learners from other backgrounds (EN-O) tend to identify different words as unintelligible.

Another interesting observation is that while the ASR-based method yields consistent results among annotators EN-J and EN-O, it does not surpass the performance of our proposed multi-task method on EN-A. This discrepancy is likely because ASR focuses on common acoustic features across accents, whereas our approach captures rater-dependent features, thereby more closely aligning with the rater’s judgment.

\section{Conclusions and Future Work}
In this study, we present a perception-based L2 speech intelligibility indicator that leverages native rater shadowing data within a sequence-to-sequence voice conversion framework. Our experiments show that this approach provides effective feedback by pinpointing unintelligible segments, offering valuable insights for CALL systems, marking the first application of such a method.

While recent work of speech assessment has focused on evaluating speech quality and naturalness \cite{saeki2022utmos, 10832295, shi2024versa}, speech intelligibility is still commonly measured using ASR's word error rates, which does not align well with real human's perception. Moreover, most existing methods provide holistic ratings over entire utterances, whereas intelligibility assessment requires a more fine-grained analysis. Shadowing, on the other hand, offers detailed insights into intelligibility and may lead to improved assessments that better accommodate the linguistic diversity of English speakers. Future work will extend our approach to cover a wider range of languages, accents, and diverse raters, facilitating broader multilingual accessibility.

\bibliographystyle{IEEEtran}
\bibliography{mybib}

\end{document}